\documentclass[12pt]{article}
\begin{document}

\newcommand{\p}{\partial}
\newcommand{\be}{\begin{equation}}
\newcommand{\ee}{\end{equation}}
\newcommand{\del}{\nabla}
\newcommand{\G}{\Gamma}
\newcommand{\g}{\sqrt{-g}}
\newcommand{\half}{\frac{1}{2}}

{\hfill   September 1997 }

{\hfill    WATPHYS-TH97/16}

\vspace*{2cm} 

\begin{center}
{\large\bf Palatini Variational Principle for $N$-Dimensional Dilaton Gravity}
\end{center}
\begin{center}
{\large\bf Howard Burton\footnote{email: hsburton@avatar.uwaterloo.ca}
and
Robert B.~Mann\footnote{e-mail: mann@avatar.uwaterloo.ca}} 
\end{center}
\begin{center}
{\it Department of Physics, University of Waterloo, Waterloo, Ontario 
N2L 3G1, Canada}  
\end{center}
\vspace*{1cm}
\begin{abstract}
\noindent
We consider a Palatini variation on a general $N$-Dimensional second 
order, torsion-free dilaton gravity action and determine the resulting
equations of motion.  Consistency is checked by considering the restraint 
imposed due to invariance of the matter action under simple coordinate
transformations, and the special case of $N=2$ is examined.  We also 
examine a sub-class of theories whereby a Palatini variation dynamically 
coincides with that of the "ordinary" Hilbert variational principle; 
in particular we examine a generalized Brans-Dicke theory and the associated
role of conformal transformations.
\end{abstract}
\newpage
\baselineskip=.8cm

\section{Introduction}

Dilaton theories of gravity are playing an increasingly important role
in the study of gravitational physics.  Such theories are a generalization
of general relativity  in which the basic field variables that describe gravity
consist of a symmetric rank-2 tensor (the metric) and one (sometimes more)
scalar field referred to as the dilaton.  The prototype of this class of 
theories is Brans-Dicke theory \cite{BD}, whose original motivation 
stemmed from developing a theory which incorporated Mach's principle by
relating the gravitational constant $G$ to the mean value of a scalar
field which was coupled to the mass density of the universe \cite{wein}.
This motivation has largely been transplanted by superstring theories \cite{str},
which generically predict that the low-energy effective Lagrangian
governing gravitational dynamics is that of a dilaton theory of gravity.

The generic form for the gravitational action for such theories is
\be\label{1}
S =  \int d^{N}x \g \: [
D(\Psi)R(g) + A(\Psi)(\del\Psi)^{2} + 16\pi\mathcal{L}_{m}(\Phi,\Psi)] 
+ S_B
\ee
\medskip
where $\Psi$ is the dilaton field, $g_{\mu\nu}$ is the metric, and $\Phi$
symbolically denotes the matter fields, whose Lagrangian ${\mathcal L}_m$
may or may not also have explicit dependence on $\Psi$.  The gravitational
field equations for such an action are derived by extremizing it with
respect to variations in the metric and dilaton fields.  Because the
first term in the action is of 2nd order in metric derivatives, it is 
necessary to add on the boundary term $S_B$ in (\ref{1}) so
that the variational principle is well defined. In particular,
both the variation of the induced metric and its derivatives must be held fixed
on the boundary.  Alternatively, the action may be supplemented with additional
boundary terms such that we need only fix the induced metric on the boundary.
Inclusion of such boundary terms is essential in order to correctly
the thermodynamics of a system of matter fields coupled to 
dilatonic gravity \cite{qgag}.

However an alternate variational principle exists for gravitational theories
in which the connection is elevated to the status of an independent gravitational
field variable, on par with the metric and the dilaton (if any).  Referred to as
the Palatini variational principle, the action for $N$-dimensional 
general relativity takes the form
\be\label{2}
S_{EH}[g,\Gamma] = \int d^{N} x \left[ \g \left( R(\Gamma) + 16\pi\mathcal{L}_{m} 
\right) \right]
\ee
where
\be\label{1a}
R(\Gamma) = g^{\mu\nu}R_{\mu\nu}(\gamma) 
= g^{\mu\nu}\left[\G^{\rho}_{\mu\nu,\rho} - \G^{\rho}_{\mu\rho,\nu} +
\G^{\rho}_{\rho\eta} \G^{\eta}_{\mu\nu} 
- \G^{\rho}_{\nu\eta} \G^{\eta}_{\mu\rho}\right]
\ee
is the Ricci scalar.  In the Palatini approach the action (\ref{2})
is subject to the independent variations $\delta_{g} S = 0$ 
and $\delta_{\Gamma} S = 0$, of the metric and the connection 
$\Gamma^\alpha_{\mu\nu}$ respectively. The former variation yields
\be\label{3}
8\pi T_{\mu\nu} = G_{\mu\nu}(\G),
\ee
\medskip
where $T_{\mu\nu}$ is the stress energy of the matter fields and $G_{\mu\nu}$
is the Einstein tensor of the manifold. Variation with respect to the
connection yields 
\be\label{4}
\partial_\lambda g_{\mu\nu} -
\G^\eta_{\lambda\mu} g_{\eta\nu} -
\G^\eta_{\lambda\nu} g_{\mu\eta}  = 0
\ee
which is the condition of metric compatibility, whose solution
\be\label{5}
\G^{\eta}_{\mu\nu} = \left\{ \!\!\!\!\!\! \begin{array}{c} \eta \\ 
\begin{array}{cc} \mu & \nu \end{array} \end{array} \!\!\!\!\!\!
\right\}
\ee
\medskip
is the Christoffel symbol.  Hence the geometrical constraint (\ref{4}) 
implicitly assumed in (\ref{1}) arises as a field equation. A curious feature
of the above approach is that there is no need to include a boundary term,
since the field variations are assumed to vanish on the boundary \cite{Wald}.

We consider in this paper the field equations and resultant dynamics which
arise from a Palatini variational principle applied to dilatonic gravitational
theories.  This ``connection-oriented" perspective is in part motivated by a potential
future quantization procedure anticipated by dilaton gravity theories, as well as
by an interest in exploring the relationship between metric compatibility and
extremization of the action.  For $N$-dimensional general relativity we have
demonstrated in a recent paper \cite{pal2} that this relationship can be 
understood to arise as a consequence of the breaking of a maximal deformation symmetry 
\cite{Hehl} associated with a transformation of the connection variables.
In this paper, however, we are solely concerned with how metricity arises (or not)
explicitly resulting from the contributions of the dilaton sector of the generalized action.
That is, we deliberately break the aforementioned deformation symmetry to isolate
dilaton-induced effects and 
chose as our starting point a generalized dilaton action whose constraints are
solely that it be first-order in curvature terms, at most quadratic in derivatives of 
$\Psi$, and with a matter action only dependent on the metric (and hence independent of both the connection 
and the dilaton field).

The action we consider is thus of the form
\begin{eqnarray}
S & = & \int d^{N}x \g \: [D(\Psi)R(\Gamma) + A(\Psi)(\del\Psi)^{2} 
+ B(\Psi)(\del^{\nu}\Psi)g_{\alpha\beta}\del_{\nu}g^{\alpha\beta} \nonumber \\
&    & \hspace{.84in} + \:C(\Psi)(\del_{\mu}\Psi)(\del_{\nu}g^{\mu\nu}) 
+ F(\Psi)\del^{2}\Psi + 16\pi\mathcal{L}_{m}(\Phi)] 
\label{6}
\end{eqnarray}
\medskip
and is clearly a function of three independent gravitational field variables:
the connection, the metric and the dilaton field. 

Note that although $\del_\mu\Psi = \p_\mu\Psi$ because $\Psi$ is a scalar, 
since metricity is not assumed, $\del^{2}\Psi$ above is given explicitly 
by $\del^{\mu}\p_{\mu}\Psi$ or $g^{\mu\nu}\del_{\nu}\p_{\mu}\Psi$.  
Clearly in an \emph{a priori} metric theory both the third and fourth terms above 
are identically zero, while the fifth merely adds a total divergence combined with a  
redefinition of the $A(\Psi)$ term.  However, if we are to take the spirit
of the Palatini variation seriously ~\cite{pal}, i.e. assume potential 
non-metricity from the outset, then these terms must occur for completeness. 

Upon investigating the dynamics resulting from a Palatini variation of the above
action, we find the circumstances under which metric compatibility explicitly 
occurs in the general $N$-dimensional case.  We find as well that the case $N=2$
merits special attention; and we investigate the differing field and geometrical 
relationships which arise in this context.  Finally, using conformal
transformations, we examine the constraints required to establish an equivalence 
between the above "Palatini dynamics" and those instead derived from the more usual
"Hilbert variational principle" - i.e. mandating {\it a priori} the equivalence 
of the connection with the Christoffel symbol and then varying solely with respect 
to the metric and the dilaton field. 

\section{$N$-Dimensional Dynamics}

If we vary the action (\ref{6}) with respect to the connection, metric and dilaton 
field respectively, we find the following field equations: 
\begin{eqnarray}
\{
\frac{-1}{\g}\del_{\lambda} \left[
D \g g^{\mu\nu} \right] + \frac{1}{2\g} \del_{\rho} \left[
D \g \left(
g^{\mu\rho}\delta^{\nu}_{\lambda} +  g^{\nu\rho}\delta^{\mu}_{\lambda}
\right) \right] &  & \nonumber \\ 
+ (B + \frac{1}{2}C) \left[
(\p^{\nu}\Psi)\delta^{\mu}_{\lambda} + (\p^{\nu}\Psi)\delta^{\nu}_{\lambda}
\right] +
(C - F)(\p_{\lambda}\Psi)g^{\mu\nu}\} & = & 0
\label{7} 
\end{eqnarray}
\begin{eqnarray}
8\pi T_{\mu\nu} & = &
DG_{(\mu\nu)} + A \left[
\p_{\mu}\Psi \p_{\nu}\Psi - \frac{1}{2}g_{\mu\nu}(\p\Psi)^{2} 
\right] - B(\p_{\mu}\Psi)\frac{\del_{\nu}\g}{\g} - B(\p_{\nu}\Psi)\frac{\del_{\mu}\g}{\g}  \nonumber \\ 
& & - B(\p^{\eta}\Psi)g_{\rho\mu}g_{\epsilon\nu}\del_{\eta}g^{\rho\epsilon} -
\frac{1}{2}Cg_{\mu\nu}(\p_{\rho}\Psi)(\del_{\eta}g^{\eta\rho}) +
F \left[
\del_{\mu}(\p_{\nu}\Psi) - \frac{1}{2}g_{\mu\nu}\del^{2}\Psi \right] \nonumber \\
& & - \del_{\rho} \left[
B \p^{\rho}\Psi g_{\mu\nu} \right] -
\frac{1}{2\g} \del_{\nu} \left[
C \g (\p_{\mu}\Psi) \right] -
\frac{1}{2\g} \del_{\mu} \left[
C \g (\p_{\nu}\Psi) \right] 
\label{8} 
\end{eqnarray}
\begin{eqnarray}\label{9} 
\{ D'R + A'(\p\Psi)^{2} - 2B'(\p^{\rho}\Psi)\frac{\del_{\rho}\g}{\g} +
C'(\p_{\mu}\Psi)\del_{\nu}g^{\mu\nu} + F'(\del^{2}\Psi) &  & \nonumber \\
+ \frac{\del_{\mu}}{\g} \del_{\nu} \left[
F \g g^{\mu\nu} \right] - \frac{\del_{\rho}}{\g} \left[ 2A \g (\p^{\rho}\Psi) \right] &  & 
 \\ 
+ \frac {2\del_{\mu}}{\g} \left[
B \g g^{\mu\nu} \left(
\frac{\del_{\nu} \g}{\g} \right) \right] - 
\frac{\del_{\mu}}{\g} \left[
C \g \del_{\nu}g^{\mu\nu} \right] \} & = & 0 \nonumber 
\end{eqnarray}
where the explicit dependence of A, B, C, D and F on $\Psi$ is suppressed 
for notational convenience; and $A'$, say, represents $\frac{\p A}{\p \Psi}$.

Consider next simplification of the connection equation (\ref{7}).
After contracting $\lambda$ and $\mu$ we find
\be\label{10} 
\frac{\del_{\rho}}{\g} \left[
D \g g^{\rho\nu} \right] = 
(\p^{\nu}\Psi) \left[
\frac{2}{1-N} \right]
\left[
(N+1)B + \left( 
\frac{N+3}{2} \right) C - F \right]
\ee
which yields upon substitution back into (\ref{7})
\be\label{11} 
(2-N)D \left(
\frac{\del_{\lambda}\g}{\g} \right) = 
\left[ \frac{(\p_{\lambda}\Psi)}{N-1} \right] \left[
N(N-1)D'+4B+(4-N^{2}+N)C+(N-2)(N+1)F \right],
\ee
after tracing over $\mu$ and $\nu$.

Therefore, assuming $N \neq 2$, we find that (\ref{7}) becomes
\be\label{12} 
\del_{\lambda} g^{\mu\nu} = X(\p_{\lambda}\Psi)g^{\mu\nu} + Y \left[
(\p^{\mu}\Psi)\delta^{\nu}_{\lambda} + (\p^{\nu}\Psi)\delta^{\mu}_{\lambda}
\right],
\ee
where (\ref{10}) and (\ref{11}) have been employed, and 
where
\be\label{13} 
X(\Psi) = \left\{
\frac{2 \left[
(1-N)D' - 2B + (N-3)C + (2-N)F \right]}{D(N-2)(1-N)}
\right\},
\ee
and
\be\label{14} 
Y(\Psi) = \left[ 
\frac{2(B+C) - F}{D(1-N)} \right] \qquad .
\ee
By permuting (\ref{12}) we can obtain an explicit solution for the connection 
\be\label{15}
\G^{\eta}_{\mu\nu} = \left\{ \!\!\!\!\!\!\! \begin{array}{c} \eta \\ 
\begin{array}{cc} \mu & \nu \end{array} \end{array} \!\!\!\!\!\!\!
\right\} + \left(
Y - \frac{1}{2}X \right)
(\p^{\eta}\Psi)g_{\mu\nu} +
\frac{1}{2}X \left[
(\p_{\mu}\Psi)\delta^{\eta}_{\nu} + (\p_{\nu}\Psi)\delta^{\eta}_{\mu} \right]
\ee
in terms of the metric and dilaton. 

{}From the form of $\G^{\eta}_{\mu\nu}$ above, 
we see that it is still symmetric 
in the lower two indices. Hence $R_{\alpha\beta}$ and $G_{\alpha\beta}$ remain symmetric 
tensors and we can replace $G_{(\alpha\beta)}$ in (3) by just $G_{\alpha\beta}$.
Furthermore, we can use the explicit form of the connection as given above to 
obtain a general expression for $G_{\mu\nu}$ in terms of the 
Christoffel symbols and the dilaton factors.  From (\ref{1a}) 
we have
\begin{eqnarray}
R_{\mu\nu} & = & R_{\mu\nu}(\{\}) + \left[
(Y'- \frac{1}{2}X') - (Y - \frac{1}{2}X) \left[ \left(
\frac{N-2}{2} \right) X - NY \right] \right] (\p \Psi)^{2} g_{\mu\nu} \nonumber \\
&   & + \left[ \left[
\frac{2-N}{2}X' - Y' \right] + \frac{4-N}{4} X^{2} - Y^{2} \right]
\p_{\mu}\Psi \p_{\nu}\Psi \nonumber \\ 
&   & + \left[
\frac{2-N}{2} X - Y \right] \del_{\mu} \p_{\nu} \Psi +
\left[
Y - \frac{1}{2}X \right]
(\del^{2}\Psi) g_{\mu\nu} 
\label{16}
\end{eqnarray}
which upon insertion into (\ref{8}) yields
\be\label{17}
8\pi T_{\mu\nu} = DG_{\mu\nu}(\{\}) + \alpha(\p\Psi)^{2}g_{\mu\nu}
+ \beta (\p_{\mu}\Psi)(\p_{\nu}\Psi) - D'[\del_{\mu}(\p_{\nu}\Psi) - (\del^{2}\Psi)g_{\mu\nu}],
\ee
where
\begin{eqnarray}
\alpha(\Psi) & = & D'' + \half F' - \half A + \half D' \left[
(3-N)X + 2(N-1)Y \right] \nonumber \\ 
&  & + \frac{1}{4} (1-N)D \left[ \left( \frac{N-2}{2} \right)
X^{2} - 2Y^{2} - (N-2)XY \right]
\label{18}
\end{eqnarray}
\be\label{19}
\beta(\Psi) = A - F' - D'' - D'X + \half (N-1) D \left[ \left(
\frac{N-2}{2} \right) X^{2} - 2Y^{2} - (N-2)XY \right].
\ee
and where (\ref{11}) has been used.

A similar substitution transforms (\ref{9}) into
\be\label{20}
D'R(\{\}) + \Pi (\p\Psi)^{2} + \Lambda (\del^{2}\Psi) = 0,
\ee
where
\begin{eqnarray}
\Pi(\Psi) & = & F'' - A' + A \left[
(N-2)X - 2NY \right] + 2F' \left[ \left(
\frac{2-N}{2} \right) X + NY \right] \nonumber \\ 
&   & + \half (1-N)D' \left[ \left(
\frac{N-2}{2} \right) X^{2} - 2Y^{2} - (N-2)XY \right] \nonumber \\ 
&   & + (1-N)D \left[ \left(
\frac{N-2}{2} \right) XX' - 2YY' - \left( \frac{N-2}{2} \right) \left[ 
XY' + YX' \right] \right], 
\label{21}
\end{eqnarray}
and
\be\label{22}
\Lambda(\Psi)  =  2F' - 2A + (1-N)D \left[ \left(
\frac{N-2}{2} \right) X^{2} - 2Y^{2} - (N-2)XY \right].
\ee
\medskip
It will later prove convenient to re-express (\ref{17}) and (\ref{20}) via (\ref{15})
directly in terms of ${\cal D}_{\mu}$, defined as the covariant derivative, $\del_{\mu}$,
with the connection equivalent to the Christoffel symbol.
In this way, then, we find that (\ref{17}) becomes:
\be
8\pi T_{\mu\nu} = DG_{\mu\nu}(\{\}) + \tilde{\alpha}(\p\Psi)^{2}g_{\mu\nu}
+ \tilde{\beta}(\p_{\mu}\Psi)(\p_{\nu}\Psi) - D' \left[ 
D_{\mu}(\p_{\nu}\Psi) - ({\cal D}^2 \Psi)g_{\mu\nu} \right],
\label{80}
\ee
\medskip
where
\be
\tilde{\alpha} := \alpha - \half D' \left[
(3-N)X + 2(N-1)Y \right],
\ee
\be
\tilde{\beta} := \beta - D'X,
\ee
and where ${\cal D}^2 \Psi$ is defined in the usual way as
\be
{\cal D}^2 \Psi := g^{\mu\nu}{\cal D}_{\mu}{\cal D}_{\nu}\Psi.
\ee
Meanwhile (\ref{20}) becomes
\be
D'R(\{\}) + \tilde{\Pi} (\p\Psi)^2 + \Lambda({\cal D}^2 \Psi) = 0,
\label{81}
\ee
where 
\begin{eqnarray}
\tilde{\Pi} & = & F'' - A' + \half (1-N)D' \left[  \left(
\frac{N-2}{2} \right) X^{2} - 2Y^{2} - (N-2)XY \right] \nonumber \\
&   & +(1-N)D[ \left( \frac{N-2}{2} \right) XX' - 2YY' -\left(
\frac{N-2}{2} \right) [XY'+YX'] \nonumber \\
&   & \hspace{0.9in} +\{ \frac{2-N}{2} X^{2}+2Y^{2}+(N-2)XY \} \{ \frac{2-N}{2} X+NY \}]
\end{eqnarray}
\medskip

As a way of checking these dynamical equations, consider the behaviour of
the matter action under an infinitesimal coordinate transformation ~\cite{wein}.  
For coordinate invariance of the matter action the condition
\be\label{23}
\p_\nu\left[\g T^{\nu}_{\lambda} \right] - \half \left(
\p_\lambda g_{\mu\nu} \right) \g T^{\mu\nu} = 0.
\ee
must hold. In the metrically compatible case, this leads directly to 
the covariant conservation of the stress energy tensor, $\del^{\eta}T_{\eta\lambda} = 0$. 
However in the more general dilatonic case, with the connection determined 
by (\ref{15}) above, we have instead the condition
\be\label{24}
\del^{\eta}T_{\eta\lambda} = -\half TX (\p_{\lambda}\Psi) +
T_{\eta\lambda} \left[ \left(
\frac{N-2}{2} \right) X - (N+1)Y \right] (\p^{\eta}\Psi) \equiv \mathcal{W}_{\lambda},
\ee
where $T$ represents the trace of the stress-energy tensor, $g^{\mu\nu}T_{\mu\nu}$.
Explicitly evaluating $\mathcal{W}_\lambda$ using (\ref{17}) yields
\begin{eqnarray}
\mathcal{W}_{\lambda} & = & \left( \frac{1}{8\pi} \right) \{
DR_{\eta\lambda}(\{\})(\p^{\eta}\Psi) [(\frac{N-2}{2}) X - Y(N+1)] \nonumber \\
&   & \hspace{.6in} -D'(\p^{\eta}\Psi) 
(\del_{\eta}(\p_{\lambda}\Psi)) [(\frac{N-2}{2}) X - Y(N+1)] \nonumber \\ 
&   & \hspace{.6in} +(\p_{\lambda}\Psi)[
\half D R(\{\}) Y(N+1) - D'( 
\half X + (N+1)Y) ({\cal D}^2 \Psi) \nonumber \\ 
&   & \hspace{1.2in} -\{X 
(\half \beta(3-N) + \alpha) +
(N+1)Y (\alpha + \beta)\} (\p\Psi)^{2}] \} 
\label{25}
\end{eqnarray}

Alternatively if we compute $\del^{\eta}T_{\eta\lambda}$ directly, by operating 
on $T_{\eta\lambda}$ as given in (\ref{17}) above with the operator $\del^{\eta}$
we obtain
\be\label{26}
\del^{\eta}T_{\eta\lambda} = \mathcal{W}_{\lambda} - \half (\p_{\lambda}\Psi) \left[
D'R(\{\}) + \tilde{\Pi}(\p\Psi)^{2} + \Lambda({\cal D}^2 \Psi)
\right],
\ee
where we have used (\ref{15}), where $\mathcal{W}_\lambda$ is given by (\ref{25}).

Hence the covariant conservation of the stress-energy is satisfied whenever 
the dilaton field equation (\ref{20}), is satisfied as well. The conservation law
(\ref{24}) generalizes to the Palatini formalism
that found in ref. \cite{qgag} for dilaton gravity theories.

\section{$N=2$ Dynamics}

We can see from the form of equation (\ref{11}) that for $N=2$ the approach
given above will break down: we will no longer be able to find an explicit expression 
for $\left( \frac{\del_{\lambda}\g}{\g} \right)$, and hence eventually 
$\del_{\lambda}g^{\mu\nu}$ in terms of functions of the dilaton field
and its derivative.  Instead, for $N=2$, we are merely left with an 
added constraint:
\be\label{27}
D' + 2B + C = 0.
\ee

Note that if (\ref{27}) does not hold then from (\ref{11}) the dilaton
must be constant $\Psi = \Psi_0$.  The field equations (\ref{7}), (\ref{8}) 
then reduce to
\be\label{28}
\frac{-1}{\g}\del_{\lambda} \left[ D_0 \g g^{\mu\nu} \right] = 0
\ee
and
\be
8\pi T_{\mu\nu} = D_0 G_{(\mu\nu)}(\Gamma)
\label{29} 
\ee
where $D_0 = D(\Psi_0)$ is constant.  This situation was previously investigated in
ref \cite{n2}.  Although it appears to yield non-trivial dynamics, this does not
occur because eq. (\ref{28}) is invariant under the transformation
\be\label{30}
\G^{\eta}_{\mu\nu} \Rightarrow \widehat{\G}^{\eta}_{\mu\nu} =
\G^{\eta}_{\mu\nu} + A_{\mu}\delta^{\eta}_{\nu} + A_{\nu}\delta^{\eta}_{\mu} -
g_{\mu\nu}A^{\eta},
\ee
where $A_\lambda$ is an arbitrary vector field. From this it may be
shown \cite{n2} that the general solution to (\ref{28}) is 
\be\label{31}
\G^{\eta}_{\mu\nu} = \left\{ \!\!\!\!\!\! \begin{array}{c} \eta \\ 
\begin{array}{cc} \mu & \nu \end{array} \end{array} \!\!\!\!\!\!
\right\} + A_{\mu}\delta^{\eta}_{\nu} + A_{\nu}\delta^{\eta}_{\mu} -
g_{\mu\nu}A^{\eta}
\ee
where $A_\mu$ is undetermined.  Insertion of this into the right hand side of
(\ref{29}) yields $G_{(\mu\nu)}(\Gamma) = 0$. Hence the theory is either 
inconsistent (if $T_{\mu\nu}\neq 0$) or trivial (if $T_{\mu\nu} = 0$). 

For $\Psi$ not constant we can understand the constraint (\ref{27}) in the 
following way. For $N=2$ the associated action
(\ref{6}) is invariant under the transformation (\ref{29})
provided the constraint (\ref{27}) is valid.  Since $A_{\lambda}$ is arbitrary, 
we can choose it in such a way as to 
achieve explicit dynamical equations for $N=2$. Since under (\ref{30})
\be\label{32}
\frac{\widehat{\del}_{\lambda} \g}{\g} = \frac{\del_{\lambda} \g}{\g} - 2A_{\lambda}.
\ee
we chose
\be\label{33}
A_{\lambda} = \half \left( \frac{\del_{\lambda}\g}{\g} \right)
\ee
so that
\be\label{34}
\del_{\lambda}g^{\mu\nu} = Y \left[ (\p^{\mu}\Psi)\delta^{\nu}_{\lambda} +
(\p^{\nu}\Psi)\delta^{\mu}_{\lambda} - (\p_{\lambda}\Psi)g^{\mu\nu} \right]
\ee
and
\be\label{35}
\G^{\eta}_{\alpha\beta} = \left\{ \!\!\!\!\!\! \begin{array}{c} \eta \\ 
\begin{array}{cc} \alpha & \beta \end{array} \end{array} \!\!\!\!\!\!
\right\} - \half Y \left[
(\p_{\alpha}\Psi)\delta^{\eta}_{\beta} + (\p_{\beta}\Psi)\delta^{\eta}_{\alpha} -
3g_{\alpha\beta}(\p^{\eta}\Psi) \right]
\ee
where the hat notation has been dropped and $B(\Psi)$ has been eliminated using
(\ref{27}).  Inserting these equations into
the field equations (\ref{8}) and (\ref{9}), together with (\ref{35}) yields
\begin{eqnarray}
8\pi T_{\mu\nu} & = &  \left[
\half D (Y' + Y^{2}) + \half D'Y(4-3N) \half A - B' - Y(C+2B) \right] (\p\Psi)^{2} g_{\mu\nu} \nonumber \\
&   & +[A - C' - D(Y' + Y^{2}) - D'Y] (\p_{\mu}\Psi)(\p_{\nu}\Psi) - 
D'[{\cal D}_{\mu}(\p_{\nu}\Psi)  - ({\cal D}^2 \Psi)g_{\mu\nu}], 
\label{36}
\end{eqnarray}
and
\begin{eqnarray}
&&\{ F''-A'+2Y'(F-C)+Y[(3N-6)A+(2-3N)F']\nonumber\\
& \qquad& +Y^{2}[7D'+6(F-C)-3N(F-C+D')] \}
(\p\Psi)^{2} \nonumber\\
&\qquad\qquad& + D'R(\{\}) + 2 \left[
F' + Y(F-C+D') - A \right]
({\cal D}^2 \Psi) =  0, 
\label{37}
\end{eqnarray}
\medskip
That is,
\be
D'R(\{\}) + \hat{\Pi}(\p\Psi)^{2} + \hat{\Lambda}({\cal D}^2 \Psi) = 0,
\ee
with the obvious definitions for $\hat{\Pi}$ and $\hat{\Lambda}$ in
accordance with (44) above.
\medskip
The conservation law (\ref{24}) holds where now
\be\label{38}
\mathcal{W}_{\lambda} := -\half TX (\p_{\lambda}\Psi) -
3T_{\eta\lambda}Y (\p^{\eta}\Psi) \quad .
\ee
However operating with $\del^{\eta}$ on both sides of (\ref{36})
leads to
\be\label{39}
\del^{\eta}T_{\eta\lambda} = \mathcal{W}_{\lambda} - \half (\p_{\lambda}\Psi) \left[
D'R(\{\}) + \hat{\Pi}(\p\Psi)^{2} + \hat{\Lambda}({\cal D}^2 \Psi) \right]
\ee
as with the $N>2$ case. Once again we see that conservation of stress energy is
consistently satisfied provided the dilaton field equation is satisfied as well.

\section{Analysis}

We turn now to a comparison of the Palatini method to the
``Hilbert variational method" - i.e. the method of varying only with respect to the metric 
and the dilaton field, assuming the metric compatibility condition (\ref{5}) is
satisfied.
In our formalism, the E-H action (\ref{2}) is equivalent to a special case of
the action (\ref{6}) with $D=1; A=B=C=F=0$,
which in turn implies, via (\ref{13}) and (\ref{14}), that $X=Y=0$, and hence
the constraint (\ref{5}), $\G = \{\}$.

Can these ideas be generalized?  Clearly from (\ref{15}), we will recover 
explicit metricity if $X=Y=0$.  From (\ref{13}) and (\ref{14}), this 
immediately implies (\ref{27}).

Yet, somewhat surprisingly, this is not the only case where the dynamics deduced from  
a Palatini variation agree with those deduced from a Hilbert variation.
Lindstr\"{o}m ~\cite{l1,l2} showed, when examining Brans-Dicke-type theories under a 
Palatini variational principle, that both the Palatini and Hilbert variations 
yield identical dynamics, the only (nominal) difference occurring in the value
of the dimensionless coupling constant $\omega$.  More specifically, under
a Palatini variation of the general action
\be\label{40}
S = \int d^{4} x \g \left[ R \psi^{\alpha} - \left( 
\omega \psi^{\alpha-2} \right)
(\p\psi)^{2} + \psi^{\beta} 16\pi\mathcal{L}_{m} \right],
\ee
he showed that the Palatini induced dynamics are equivalent to the
Hilbert ones, with a rescaling of the (dimensionless) coupling constant:
\be\label{41}
\omega \rightarrow \widehat{\omega} = \omega - \frac{3\alpha^{2}}{2}.
\ee
The justification for this equivalence lies
in the fact that for this particular action the form of the connection 
constraint (\ref{15}) is
\be\label{42}
\G^{\eta}_{\mu\nu} = \left\{ \!\!\!\!\!\! \begin{array}{c} \eta \\ 
\begin{array}{cc} \mu & \nu \end{array} \end{array} \!\!\!\!\!\!
\right\} + \left( \frac{\alpha}{2\psi} \right) \left[
(\p_{\mu}\psi) \delta^{\eta}_{\nu} + (\p_{\nu}\psi) \delta^{\eta}_{\mu} -
g_{\mu\nu}(\p^{\eta}\psi) \right],
\ee
which we recognize as that of an induced metrically-compatible connection after
a conformal transformation
\be\label{43}
g_{\mu\nu} \rightarrow \widehat{g}_{\mu\nu} = \psi^{\alpha} g_{\mu\nu}; 
\psi \rightarrow \widehat{\psi} = \psi,
\ee
\medskip
expressed in terms of the ``old" metric $g_{\mu\nu}$.  Therefore if we
apply the following conformal transformation, henceforth known as a "Palatini Transformation"
(owing to its explicit mention of the connection) 
\be\label{44}
g_{\mu\nu} \rightarrow \widehat{g}_{\mu\nu} = \psi^{\alpha} g_{\mu\nu};
\psi \rightarrow \widehat{\psi} = \psi;
\G^{\eta}_{\rho\tau} \rightarrow \widehat{\G}^{\eta}_{\rho\tau} = 
\G^{\eta}_{\rho\tau},
\ee
to the action (\ref{40}), we find  
\be\label{45}
\widehat{S} = \int d^{4} x \sqrt{-\widehat{g}} \left[
\widehat{R} - \left( 
\frac{\omega}{\psi^{2}} \right)
\widehat{(\p\psi)}^{2} + \psi^{\beta-2\alpha} \mathcal{L}_{m} \right],
\ee
which can now be regarded as an Einstein-Hilbert action coupled to matter.
The Palatini variation acting upon the transformed action (\ref{45}) gives simply
the Christoffel relation (\ref{5}), and thus for this action (\ref{45}) both
the Hilbert and Palatini variational methods lead to identical results for 
the connection.  To obtain the dynamics of (\ref{40}), then, one can apply
the Palatini variational principle to (\ref{45}) and then subject the results
to the "inverse Palatini transformation":
\be
g_{\mu\nu} \rightarrow \widehat{g}_{\mu\nu} = \psi^{-\alpha}g_{\mu\nu};
\psi \rightarrow \widehat{\psi} = \psi;
\G^{\eta}_{\rho\tau} \rightarrow \widehat{\G}^{\eta}_{\rho\tau} = \G^{\eta}_{\rho\tau}
\ee
\medskip
Similarly, if one starts again from the action (\ref{45}) and assumes, under a 
"Hilbert perspective" that the connection is always equivalent to the 
Christoffel symbol, it can be easily shown that (\ref{45}) can be transformed
into 
\be
S = \int d^{4}x \g \left[
R\psi^{\alpha} - \left(
\widehat{\omega}\psi^{\alpha-2} \right)
(\p\psi)^{2} + \psi^{\beta}16\pi \mathcal{L}_m \right],
\ee
i.e. (\ref{40}) with $\widehat{\omega}$, as given by (\ref{41}), instead of  
$\omega$.
\medskip
This treatment by Lindstr\"{o}m linking the Palatini and Hilbert approaches 
suggests that we look at an analagous 
form of our Palatini connection equation, i.e. when the connection in
(\ref{15}) has the form of a Christoffel symbol after an associated conformal transformation.
Inspection of (\ref{15}) indicates this happens if $Y=0 \Leftrightarrow 2B+2C-F = 0$.
The connection equation then becomes
\be\label{46}
\G^{\eta}_{\mu\nu} = \left\{ \!\!\!\!\!\! \begin{array}{c} \eta \\ 
\begin{array}{cc} \mu & \nu \end{array} \end{array} \!\!\!\!\!\!
\right\} + \frac{1}{2}X \left[
(\p_{\mu}\Psi)\delta^{\eta}_{\nu} + (\p_{\nu}\Psi)\delta^{\eta}_{\mu} -
(\p^{\eta}\Psi)g_{\mu\nu} \right]
\ee
which is identical to the form of the induced Christoffel connection after
a conformal transformation of the metric:
\be\label{47}
g_{\mu\nu} \rightarrow \widehat{g}_{\mu\nu} = \Omega^{2}g_{\mu\nu},
\ee
where
\be\label{48}
\p_{\epsilon}(ln \Omega) = \frac{X}{2} \p_{\epsilon}\Psi.
\ee
Therefore, if we apply the following Palatini transformation to the action (\ref{6})
\footnote{recall that here the Ricci tensor, $R_{\mu\nu}$ is expressed {\it solely}
as a function of the connection and is hence unchanged by the following transformation}:
\be\label{49}
\G^{\eta}_{\rho\xi} \rightarrow \widehat{\G}^{\eta}_{\rho\xi} = \G^{\eta}_{\rho\xi};
\Psi \rightarrow \widehat{\Psi} = \Psi; g_{\mu\nu} \rightarrow \Omega^{2}\widehat{g}_{\mu\nu}
\ee
we find the transformed action
\begin{eqnarray}
\widehat{S} & = & \int d^{N} x \sqrt{-\widehat{g}} \{
\Omega^{2-N} [D\widehat{R} + A (\widehat{\p\Psi})^{2} + B\widehat{(\p^{\eta}\Psi)} \widehat{g}_{\mu\nu}
\widehat{\del}_{\eta} \widehat{g}^{\mu\nu} + C (\p_{\mu}\Psi)(\widehat{\del}_{\nu} \widehat{g}^{\mu\nu}) \nonumber \\
&   & \hspace{1.2in} + F \widehat{\del}^{\eta}(\p_{\eta}\Psi)] + 16\pi \widehat{\mathcal{L}}_{m} +
2(NB+C)\Omega^{1-N}(\p_{\mu}\Psi)(\p_{\nu}\Omega)\widehat{g}^{\mu\nu}\} 
\label{50}
\end{eqnarray}
Applying the Palatini principle to this action together with the above  
constraints, $Y=0 \Leftrightarrow 2B+2C-F = 0$ and eq. (\ref{48})
results in
the Hilbert constraint (\ref{5}) for the connection equation, i.e. $\G = \{\}$.
This may be seen as follows.
Variation of the above action with respect to $\Gamma^\lambda_{\mu\nu}$ yields 
after some simplification the equation
\be\label{50a}
\Omega^{2-N}\del_{\alpha}\psi g^{\mu\nu}[C-F-D'] - \frac{D}{\g}\del_{\alpha} \left[
\Omega^{2-N} \g g^{\mu\nu} \right] = 0
\ee
where the constraint $Y=0$ has been imposed. Note that for $Y=0$
\be
\frac{X}{2} = \left[ \frac{D'-C+F}{D(N-2)} \right]
\ee
which implies from (\ref{50a})
\be\label{50b}
\frac{1}{\g}\del_{\alpha} \left[ \g g^{\mu\nu} \right] = 0
\ee
where the above constraint (\ref{48}) has been employed. For $N\neq 2$
it is straightforward to show that (\ref{50b}) implies that
metricity holds.  The $N=2$ case follows from the same connection invariance arguments
given in the previous section.

Unfortunately, the above analysis alone is not very helpful in answering
the question of when the Hilbert and Palatini approach differ in regards
to their \emph{physical} content.  Lindstr\"{o}m was working with a particular
action which had the unique property of being, after conformally transformed via the 
associated Palatini transformation and "re-transformed" via the associated
Hilbert transformation, equivalent to the original action with the sole exception 
that $\omega \rightarrow \hat{\omega} = \omega - \frac{3(\alpha)^{2}}{2}$.  

In general this 
is definitely not the case, i.e. one cannot generally say that the 
inverse Hilbert transformation:
\be
\widehat{g}_{\mu\nu} \rightarrow g_{\mu\nu} = \Omega^{-2}\widehat{g}_{\mu\nu};
\widehat{\Psi} \rightarrow \Psi = \widehat{\Psi}
\ee
\medskip
applied to (\ref{45}) will yield some other Hilbert action with merely a change in some 
dimensionless constant.  Nonetheless, one can say something.  To do this, we 
again regard our Palatini dynamical equations expressed explicitly in terms 
of the Christoffel symbol, i.e. (\ref{80}) and (\ref{81}), but
this time subject to the explicit constraint Y = 0.
Therefore our Palatini dynamical equations (\ref{80}),(\ref{81}) for $Y=0$ now
become:
\begin{eqnarray} 
8\pi T_{\mu\nu} & = & DG_{\mu\nu}(\{\}) + \left[
D'' + \frac{1}{2} \left(
F'-A+\tilde{Q} \right) \right] (\p\Psi)^{2} g_{\mu\nu} \nonumber \\
&    & - [D''+(F'-A+\tilde{Q})] (\p_{\mu}\Psi)(\p_{\nu}\Psi) -
D'[{\cal D}_{\mu}(\p_{\nu}\Psi) - {\cal D}^2\Psi], 
\end{eqnarray}
\medskip
and
\be
D'R(\{\}) + \left[ (F'-A+\tilde{Q})' \right] (\p\Psi)^{2} +
2 \left[ F'-A+\tilde{Q} \right] {\cal D}^2\Psi = 0,
\ee
where 
\be
\tilde{Q} := (N-1)\frac{2-N}{4}DX^{2} = \frac{(N-1)(D'-C+F)^{2}}{D(2-N)}.
\ee
\medskip
Meanwhile a Hilbert variation of our original action (\ref{6}) yields:
\begin{eqnarray} 
8\pi T_{\mu\nu} & = & DG_{\mu\nu} + \left[
D'' + \frac{1}{2} \left(
F'- A \right) \right] (\p\Psi)^{2} g_{\mu\nu} \nonumber \\
&    & - [D''+(F'-A)] (\p_{\mu}\Psi)(\p_{\nu}\Psi) -
D'[{\cal D}_{\mu}(\p_{\nu}\Psi) - {\cal D}^2\Psi], 
\end{eqnarray}
and
\be
D'R(\{\}) + \left[ (F'- A)' \right] (\p\Psi)^{2} +
2 \left[ F'- A \right] {\cal D}^2\Psi = 0,
\ee
\medskip
Therefore, provided that $\tilde{Q}$ is proportional to $(F'-A)$, the Palatini
dynamics will merely result in a rescaling of the dimensionless function $F'-A$ 
and so will be equivalent to the Hilbert dynamics, as in the above special case
examined by Lindstr\"{o}m.  Note that $\tilde{Q}$ is a quadratic function
of $D'$, $C$ and $F$, and so this allows for a more general 
set of relationships between the dilatonic functions.  

In order to illustrate the point, let us take, for example, the following
case (for $a,b,c,d,f \in \mathcal{R}$):
\be 
A = a e^{k\Psi}; D = d e^{k\Psi}; C = c e^{k\Psi}; F = f e^{k\Psi} 
\ee
\medskip
and hence, by $Y=0$, 
\be
B = \frac{f-2c}{2} e^{k\Psi}
\ee.
\medskip
Therefore, we now have:
\be
\tilde{Q} = \frac{(N-1)(kd-c+f)^{2}}{(2-N)d} e^{k\Psi},
\ee
\medskip
and clearly we now have a case where the Palatini dynamics
are physically equivalent to the Hilbert dynamics, the only difference
being, as before for the Brans-Dicke-like theories, that 
the (physically irrelevant) dimensionless constant, 
$\omega := kf-a$, now becomes:
\be
\widehat{\omega} = \omega + \frac{(N-1)(kd-c+f)^{2}}{(2-N)d} 
\ee
\bigskip
\section{Conclusions}
\medskip
We have examined the explicit dynamics of a general second-order, $N-$dimensional,
torsion-free dilaton gravity action under the Palatini variational principle 
and checked these dynamics by considering the invariance of the matter action
under simple coordinate transformations.  Unlike general relativity, derived
from the standard Einstein-Hilbert action, a Palatini variation of the action
does not generally lead to equivalent dynamics to that of a Hilbert variation
(i.e. the dynamics obtained by assuming \emph{a priori} that the connection in 
the action is that of the Christoffel symbol). Instead the dynamics are
only identical provided that $D'+2B+C=0$.  Furthermore, when both $2B+2C-F=0$, and 
$\tilde{Q} = \frac{(N-1)}{D(N-2)} (D'-C+F)^{2}$ is proportional to
$(F'-A)$, then the two approaches merely differ by a
physically irrelevant constant.

\section*{Acknowledgements}
This work was supported in part by the Natural Sciences and Engineering
Research Council of Canada.

\end{document}